\documentclass[prl,twocolumn,showpacs,superscriptaddress,nofootinbib,floatfix,showkeys,10pt]{revtex4-1}

\usepackage{graphicx}
\usepackage{amsmath}
\usepackage{bm}
\usepackage{yhmath}
\usepackage{mathtools}
\usepackage{wasysym}
\usepackage[colorlinks,citecolor=blue,urlcolor=blue,linkcolor=blue]{hyperref}
\usepackage{subfigure}
\usepackage{color}
\usepackage{cases}
\usepackage{subfigure}
\usepackage{times}
\usepackage{dcolumn,booktabs,bm}
\usepackage{slashed}
\usepackage{amsfonts,amssymb,stmaryrd,latexsym}
\usepackage{mathrsfs}
\usepackage{textcomp}
\usepackage{multirow}
\usepackage{cancel}
\usepackage{array}
\usepackage{enumerate,enumitem}
\usepackage{orcidlink}
\usepackage{dashrule}
\usepackage{multirow}
\usepackage{times}
\usepackage[normalem]{ulem}

\allowdisplaybreaks[4]
\newcounter{cnt}
\makeatletter
\let\oldhypertarget\hypertarget
\renewcommand{\hypertarget}[2]{%
  \oldhypertarget{#1}{#2}%
    \protected@write\@mainaux{}{%
        \string\expandafter\string\gdef
          \string\csname\string\detokenize{#1}\string\endcsname{#2}%
    }%
  }
\newcommand{\myhyperlink}[1]{%
  \hyperlink{#1}{\csname #1\endcsname}%
  }
\makeatother

\def\N2LO{{N$^2$LO}}

\newcommand{\rub}{\affiliation{Institut f\"ur Theoretische Physik II, Ruhr-Universit\"at Bochum, D-44780 Bochum, Germany }}
\newcommand{\pkusp}{\affiliation{School of Physics, Peking University, Beijing 100871, China}}
\newcommand{\pku}{\affiliation{School of Physics and Center of High Energy Physics, Peking University, Beijing 100871, China}}
\newcommand{\cqu}{\affiliation{Department of Physics and Chongqing Key Laboratory for Strongly Coupled Physics, Chongqing University, Chongqing 401331, China}}
\newcommand{\upc}{\affiliation{College of Science, China University of Petroleum, Qingdao, Shandong 266580, China}}
\newcommand{\xju}{\affiliation{School of Physics, Xi‘an Jiaotong University, Xi‘an 710049, China}}
\newcommand{\fdu}{\affiliation{Key Laboratory of Nuclear Physics and Ion-beam Application (MOE) and Institute of Modern Physics, Fudan University, Shanghai 200443, China}}
\newcommand{\zzu}{\affiliation{School of Physics, Zhengzhou University, Zhengzhou 450001, China}}
\newcommand{\seu}{\affiliation{School of Physics, Southeast University, Nanjing 211189, China}}

\begin{document}

%%%%%%%%%%%%%%%%open the reply mode%%%%%%%%%%%%%%%%%%
%\onecolumngrid
%\input{reply_b.tex}
%\newpage
%\setcounter{page}{0}
%%%%%%%%%%%%%%%%open the reply mode%%%%%%%%%%%%%%%%%%

\title{Emergence of the exotic bottomoniumlike  state $Y(10650)$ and support from Belle and Belle II data}

\author{Jun-Zhang Wang\,\orcidlink{0000-0002-3404-8569}}\email{wangjzh@cqu.edu.cn}\cqu\pku

\author{Yu-Bo Li\,\orcidlink{0000-0002-9909-2851}}\email{yubo.li@xjtu.edu.cn}\xju

\author{Jian-Bo Cheng\orcidlink{0000-0003-1202-4344}}\email{jbcheng@upc.edu.cn}\upc

\author{Zi-Yang Lin\,\orcidlink{0000-0001-7887-9391}}\email{lzy\_15@pku.edu.cn}\seu\pkusp

\author{Lu Meng\, \orcidlink{0000-0001-9791-7138}}\email{lmeng@seu.edu.cn}\seu\rub

\author{Cheng-Ping Shen\, \orcidlink{0000-0002-9012-4618}}\email{shencp@fudan.edu.cn}\fdu\zzu

\author{Shi-Lin Zhu\,\orcidlink{0000-0002-4055-6906}}\email{zhusl@pku.edu.cn}\pku

\begin{abstract}
Near-threshold exotic hadrons are usually associated with $S$-wave hadron-hadron dynamics, while higher partial waves are expected to be strongly suppressed by the centrifugal barrier. We show that this expectation can be overturned in the bottomonium sector. In a coupled-channel meson exchange framework combined with the complex scaling method, we find a $J^{PC}=1^{--}$ pole, denoted as $Y(10650)$, generated dominantly by the $P$-wave $B^*\bar B^*$ interaction and located close to the $B^*\bar B^*$ threshold. This pole naturally accounts for the anomalous enhancement observed just above the opening of the $B^*\bar B^*$ threshold in $e^+e^-\to B^*\bar B^*$. Once its production strength is fixed by this threshold enhancement, the corresponding cross sections of $\sigma[e^+e^-\to Y(10650)\to B\bar B^*]$ are predicted by the pole residues and phase-space factors, giving a characteristic dip-or-peak structure consistent with the available Belle~(II) data. We further study the hidden-bottom transition $Y(10650)\to \Upsilon(2S)\eta$ through a near-threshold $B^*\bar B^*$ loop mechanism. The resulting $\mathcal{O}(10\sim100~\mathrm{keV})$ width for $Y(10650)\to \Upsilon(2S)\eta$ is sufficient to account for the corresponding cross sections measured by Belle II. The simultaneous appearance of this state in open- and hidden-bottom channels provides a direct experimental path to test a $P$-wave near-threshold mechanism and makes $Y(10650)$ a strong candidate for the first neutral isoscalar exotic bottomoniumlike state in the spectral gap between $\Upsilon(4S)$ and $\Upsilon(5S)$.
\end{abstract}

%\pacs{12.39.Fe, 12.39.Hg, 14.40.Nd, 14.40.Rt}
\maketitle

\textit{Introduction.---} 
%\preivous version:
%The discovery of the $XYZ$ states has transformed heavy-quark spectroscopy from a classification of conventional $q\bar q$ mesons into a search for new forms of hadronic matter. 
The discovery of the $XYZ$ states has extended heavy-quark spectroscopy from a classification of conventional $q\bar q$ mesons to the search for new forms of hadronic matter. A large number of charmoniumlike candidates, including the $X(3872)$, the charged $Z_c$ states, and the vector $Y$ structures, have revealed the importance of multiquark dynamics, hadronic molecules, and threshold-induced coupled-channel phenomena near open-flavor thresholds~\cite{Chen:2016qju,Hosaka:2016pey,Lebed:2016hpi,Esposito:2016noz,Olsen:2017bmm,Guo:2017jvc,Liu:2019zoy,Brambilla:2019esw,Meng:2022ozq,Liu:2024uxn,Bai:2026atm,Dai:2026fkg,Workman:2022ynf}. In the bottomonium sector, however, both theoretical and experimental studies remain far less advanced. Apart from the charged $Z_b(10610)$ and $Z_b(10650)$~\cite{Belle:2011aa}, no neutral isoscalar manifestly exotic bottomoniumlike state has been established. Identifying such a state would fill a major gap in heavy-quark spectroscopy and provide a new benchmark for nonperturbative QCD in the heavy-quark limit.

A particularly clean opportunity arises in the mass region between $\Upsilon(4S)$ and $\Upsilon(5S)$. This interval is restrictive for conventional bottomonium: no additional $S$-wave vector $b\bar b$ meson is expected there, and the $\Upsilon(10750)$ observed by Belle has been proposed as a state dominated by the $3D$ bottomonium component~\cite{Godfrey:2015dia,Wang:2018rjg,Chen:2019uzm,Liang:2019geg,Li:2021jjt,Bai:2022cfz}. Remarkably, the $B^*\bar{B}^*$ threshold also falls within the same energy region. Belle~(II) measurements of $e^+e^-\to B^*\bar B^*$ cross sections show a pronounced enhancement immediately above this threshold~\cite{Belle:2021lzm,Belle-II:2024niz}. Such a rapid rise is difficult to be attributed solely to a smooth continuum contribution, as the near-threshold $P$-wave phase space is strongly suppressed. Therefore, it calls for an additional near-threshold dynamical mechanism.

Most near-threshold structures observed in heavy-quark systems have been connected to $S$-wave hadron-hadron interactions~\cite{Chen:2016qju,Lebed:2016hpi,Esposito:2016noz,Olsen:2017bmm,Guo:2017jvc,Liu:2019zoy,Brambilla:2019esw,Meng:2022ozq,Liu:2024uxn,Bai:2026atm,Dai:2026fkg,Workman:2022ynf}. This is natural: the centrifugal barrier makes higher partial waves less efficient in generating visible near-threshold poles. Nevertheless, recent studies in the charm sector have suggested that this expectation may not be universal. The structure known as $G(3900)$~\cite{BaBar:2006qlj,BaBar:2008drv,Belle:2007qxm}, recently confirmed with high precision in $e^+e^-\to D\bar D$ process~\cite{BESIII:2024ths}, has been interpreted as a possible $P$-wave dimeson resonance~\cite{Lin:2024qcq,Nakamura:2023obk,Ye:2025ywy}, and similar ideas have been discussed for other charm-strange sector enhancements~\cite{Wang:2024ukc}. These structures are intriguing, but their nature remains inconclusive, because their broad line shapes can be difficult to distinguish from nonpole mechanisms, such as coupled-channel interference effects~\cite{Eichten:1978tg}. The bottom sector offers a cleaner setting: larger meson masses reduce the kinetic energy near threshold, allowing the same interaction mechanism to generate poles closer to threshold and with narrower and more experimentally identifiable line shapes.

In this Letter, we %show that such a mechanism is realized in the open-bottom system. 
employ a coupled-channel meson exchange interaction together with the complex scaling method and find a $J^{PC}=1^{--}$ pole, denoted as $Y(10650)$, located very close to the $B^*\bar B^*$ threshold. The pole is dominantly generated by the $P$-wave $B^*\bar B^*$ interaction, with the relevant coupled channels $B^*\bar B+c.c.(^3P_1)$, $B^*\bar B^*(^1P_1)$, and $B^*\bar B^*(^5P_1)$. Its position in the spectral gap between $\Upsilon(4S)$ and $\Upsilon(5S)$, together with its near-threshold open-bottom origin, makes it a natural candidate for a neutral isoscalar exotic bottomoniumlike state.

The predicted pole is directly tied to existing data. It provides a natural explanation for the pronounced enhancement observed just above the $B^*\bar B^*$ threshold in $\sigma(e^+e^-\to B^*\bar B^*)$ distribution~\cite{Belle-II:2024niz}. Once the production strength is fixed in this channel, the absolute cross sections for $e^+e^-\to Y(10650)\to B\bar B^*$ are predicted by the pole residues and phase-space factors, rather than by an independent free adjustment. The resulting structure is consistent with the available Belle~(II) data and gives a correlated open-bottom signature of the same pole. We further show that the near-threshold $B^*\bar B^*$ component manifests itself in the hidden-bottom transition $Y(10650)\to \Upsilon(2S)\eta$, providing mutual evidence for the interpretation. In particular, the predicted partial width for $Y(10650)\to \Upsilon(2S)\eta$ is on the order of $(10\sim100~\mathrm{keV})$, sufficient to account for the cross sections recently measured by Belle II~\cite{Belle-II:2025ubm}. These open- and hidden-bottom observations promote the $Y(10650)$ from a model prediction to an experimentally testable candidate for the first neutral isoscalar exotic bottomoniumlike state.

\textit{Coupled-channel dynamics and pole prediction.---}We describe the $I=0$, $J^{PC}=1^{--}$ open-bottom dynamics in a coupled-channel $B^{(*)}\bar B^{(*)}$ system. The interaction kernel is constructed from the heavy-meson effective Lagrangians through the exchange of light pseudoscalar, scalar, and vector mesons. For a transition from channel $\beta$ to channel $\alpha$, the Born-level momentum-space potential has the generic form
\begin{eqnarray}
{\cal V}_{\alpha\beta}({\bm p},{\bm p}')
=
\sum_{\varphi=\pi,\eta,\sigma,\rho,\omega}
C_{\alpha\beta}^{\varphi}
\frac{{\cal O}_{\alpha\beta}^{\varphi}({\bm p},{\bm p}')}
{({\bm p}-{\bm p}')^2+m_\varphi^2}
{\cal F}(p^2,p^{\prime 2}),
\label{eq:born}
\end{eqnarray}
where $C_{\alpha\beta}^{\varphi}$ is the flavor factor and ${\cal O}_{\alpha\beta}^{\varphi}$ denotes the spin-momentum operator generated by the corresponding effective vertices. The explicit Lagrangians, operator structures, and coupling constants are given in the Supplemental Material (SM)~\cite{suppl}. The potential entering the dynamical equation is obtained by transforming the helicity potential into the $LSJ$ basis and then projecting onto a definite total angular momentum. Schematically,
\begin{eqnarray}
V^{J}_{L'S',LS}(p,p')
&=&
\frac{\sqrt{(2L'+1)(2L+1)}}{2J+1}
\sum_{\lambda',\lambda}
C_{L'0,S'\lambda'}^{J\lambda'} \nonumber\\
&&\times
C_{L0,S\lambda}^{J\lambda}V^{J}_{\lambda'\lambda}(p,p'), 
\label{eq:pwproj}
\end{eqnarray}
with
\begin{eqnarray}
V^{J}_{\lambda'\lambda}(p,p')
=2\pi\int_{-1}^{1}d(\cos\theta)\,
d^{J}_{\lambda\lambda'}(\theta)
{\cal V}_{\lambda'\lambda}({\bm p},{\bm p}'). 
\end{eqnarray}
Thus, the calculation is a coupled-channel matrix problem. In the $1^{--}$ sector, channels $1,2,3$ are assigned to $B^*\bar B+c.c.(^3P_1)$, $B^*\bar B^*(^1P_1)$, and $B^*\bar B^*(^5P_1)$, respectively. The $B^*\bar B^*(^3P_1)$ configuration has $J^{PC}=1^{-+}$ and is not included. In this basis,
\begin{eqnarray}
{\bf V}^{J=1}(p,p')=
\begin{pmatrix}
V_{11} & V_{12} & V_{13}\\
V_{21} & V_{22} & V_{23}\\
V_{31} & V_{32} & V_{33}
\end{pmatrix},
\label{eq:vmat}
\end{eqnarray}
where each matrix element contains the allowed $\pi,\eta,\sigma,\rho,\omega$ exchange contributions after the partial-wave projection.

The pole positions are obtained from the coupled-channel Schr\"odinger equation in momentum space.
\iffalse
\begin{eqnarray}
E\phi_\alpha({\bm p})
&=&
\left(m_{\rm th}^{\alpha}+\frac{{\bm p}^2}{2\mu_\alpha}\right) \phi_\alpha({\bm p}) \nonumber\\
&&
+ \sum_{\beta} \int\frac{d^3{\bm k}}{(2\pi)^3}
V_{\alpha\beta}({\bm p},{\bm k}) \phi_\beta({\bm k}),
\label{eq:ccsch}
\end{eqnarray}
\fi
The short-distance part of the one-boson exchange (OBE) kernel is regularized by the nonlocal monopole form factor
\begin{eqnarray}
{\cal F}(p^2,p^{\prime2})=
\frac{\Lambda^2}{\Lambda^2+p^2}
\frac{\Lambda^2}{\Lambda^2+p^{\prime2}},
\label{eq:reg}
\end{eqnarray}
with $\Lambda$ the cutoff parameter in the scattering equation. We then apply the complex scaling method~\cite{Aguilar:1971ve,Balslev:1971vb,Lin:2022wmj,Lin:2023ihj}, $U(\theta){\bm r}={\bm r}e^{i\theta}$ and $U(\theta){\bm p}={\bm p}e^{-i\theta}$, under which the coupled-channel Schr\"odinger equation becomes
\begin{eqnarray}
E\phi_\alpha(\tilde{\bm p})
&=&
\left(m_{\rm th}^{\alpha}
+\frac{{\bm p}^2e^{-2i\theta}}{2\mu_\alpha}\right)
\phi_\alpha(\tilde{\bm p})
+
\sum_{\beta}
\int\frac{d^3{\bm k}\,e^{-3i\theta}}{(2\pi)^3} \nonumber\\
&&\times
V_{\alpha\beta}(\tilde{\bm p},\tilde{\bm k})
\phi_\beta(\tilde{\bm k}),
\label{eq:ccschcsm}
\end{eqnarray}
where $m_{\rm th}^{\alpha}$ and $\mu_\alpha$ are the threshold and reduced mass of channel $\alpha$, respectively. A genuine pole in the unphysical sheet appears as a $\theta$-stable eigenvalue once $2\theta>|\mathrm{Arg}(E_R)|$ for $E_R=M_R-i\Gamma_R/2$.

\begin{figure}[ht]
\begin{centering}
\scalebox{1.00}{\includegraphics[width=\linewidth]{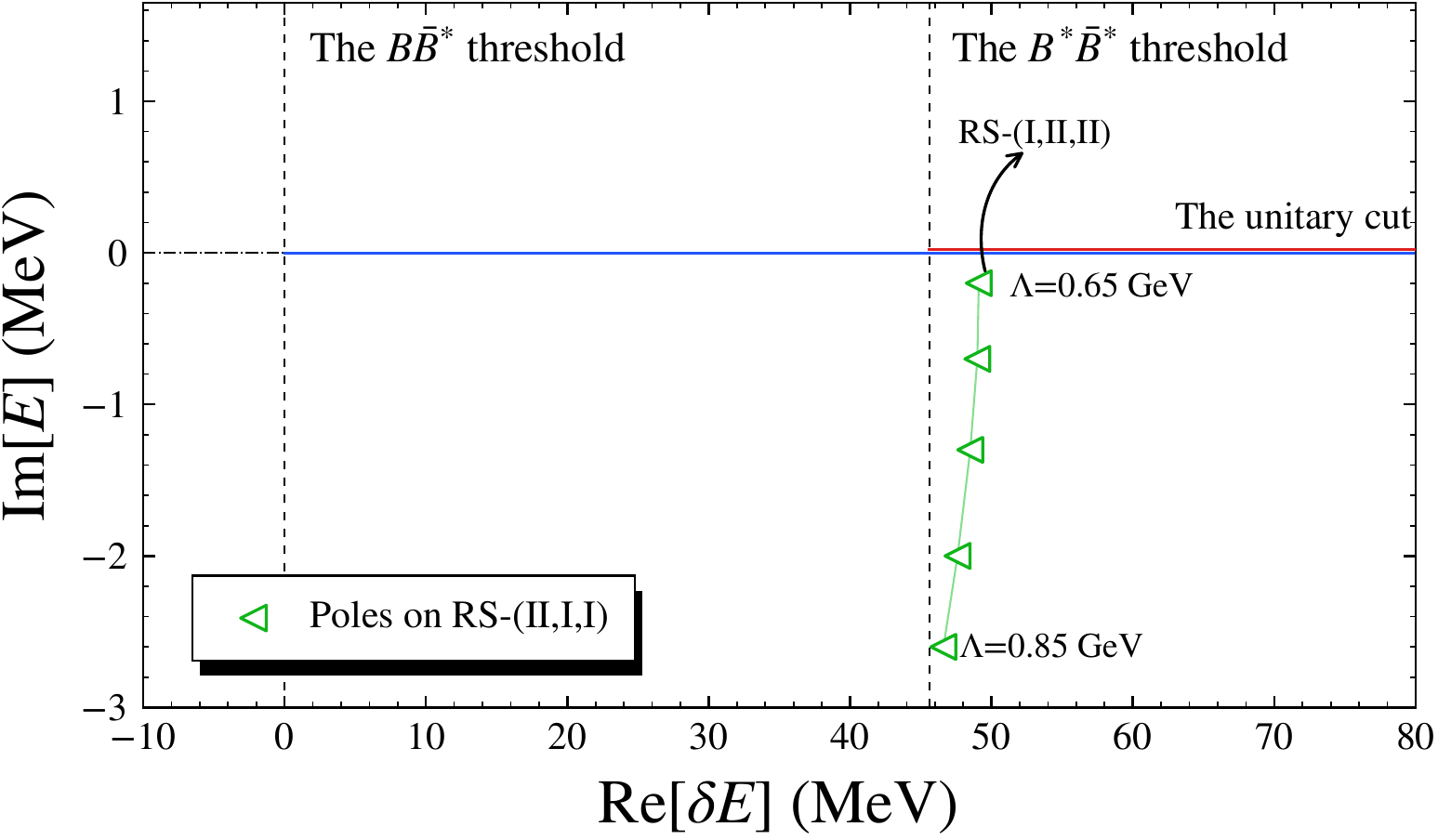}}
\caption{Pole trajectory of the predicted $Y(10650)$ in the complex energy plane. The green triangles denote the poles on RS-$(\mathrm{II},\mathrm{I},\mathrm{I})$ obtained by varying the cutoff $\Lambda$. 
%The blue and red horizontal lines indicate the unitary cuts associated with the $B\bar B^{*}$ and $B^{*}\bar B^{*}$ thresholds, respectively. 
Here $\delta E=E_{\mathrm{pole}}-m_{\rm th}^{0}$ with $m_{\rm th}^{0}=m_{B}+m_{B^*}$.}
\label{fig:poles}
\end{centering}
\end{figure}

The resulting pole trajectory is shown in Fig.~\ref{fig:poles}. In the three-channel problem, the Riemann sheet (RS) is specified by the signs of the imaginary parts of the three channel momenta across their unitary cuts. We denote them by RS-$(s_1,s_2,s_3)$, where $s_i=\mathrm{I}$ or $\mathrm{II}$ labels the physical or unphysical sheet of channel $i$, respectively. The pole found here persists on RS-$(\mathrm{II},\mathrm{I},\mathrm{I})$ for $\Lambda=0.65$--$0.85$ GeV. Its location is highly nontrivial: although $\mathrm{Re}\,[E_R]$ lies slightly above the $B^*\bar B^*$ threshold, the pole remains on the physical sheets of the two $B^*\bar B^*$ channels. It is therefore neither a subthreshold bound state nor a standard resonance pole, and its imaginary part should not be directly read as the half width of a Breit--Wigner resonance. It instead manifests itself as a threshold cusp-like structure in the line shape.  Such a pole topology has rarely appeared in previous studies of exotic hadrons~\cite{Hosaka:2016pey,Guo:2017jvc,Yamaguchi:2014era,Nieves:2012tt,Liu:2019stu,Dias:2014pva,Ohkoda:2012hv,Liu:2019tjn,Yamaguchi:2019seo,Wu:2010jy,Braaten:2004rn,Giachino:2022pws,Nakamura:2021dix,Meng:2023for,Suenaga:2021qri} and highlights a special feature of $P$-wave coupled-channel dynamics.
As $\Lambda$ varies, the pole stays close to the $B^*\bar B^*$ threshold and moves only from $10651.0-2.6i$ MeV to $10653.5-0.2i$ MeV. With decreasing attraction, the trajectory tends toward RS-$(\mathrm{I},\mathrm{II},\mathrm{II})$, as indicated schematically in Fig.~\ref{fig:poles}. 
%The corresponding pole cannot be solved in the present complex-scaling calculation because the required analytic continuation is numerically challenged. 
The central conclusion is that the $P$-wave $B^{(*)}\bar B^{(*)}$ coupled-channel interaction generates a stable threshold-pinned pole, which we identify as $Y(10650)$.

\begin{table*}[htb]
\renewcommand{\arraystretch}{1.45}
\caption{Pole properties of the predicted $Y(10650)$ for representative cutoffs. %Here $\delta E=E_{\mathrm{pole}}-m_{\rm th}^{0}$ with $m_{\rm th}^{0}=m_{B}+m_{B^*}$. 
The complex quantities $\mathcal P_i=(\phi_i|\phi_i)$ provide diagnostic channel weights in the c-product normalization, and ``Res'' denotes the pole residue coupled to each channel.}
\label{tab:10650}
\setlength{\tabcolsep}{3.0mm}
\centering
\begin{tabular}{cccccc}
\toprule[0.8pt]
States/$\Lambda$ (GeV) & 0.85 & 0.80 & 0.75 & 0.70 & 0.65 \\
\hline
Pole position (MeV) & $10651.0-2.6i$ & $10652.0-2.0i$ & $10652.9-1.3i$ & $10653.4-0.7i$ & $10653.5-0.2i$ \\
\hline
$\delta E$ (MeV) & $1.0-2.6i$ & $2.0-2.0i$ & $2.9-1.3i$ & $3.4-0.7i$ & $3.5-0.2i$ \\
\hline
Riemann sheet & (II,I,I) & (II,I,I) & (II,I,I) & (II,I,I) & (II,I,I) \\
\hline
$\mathcal P(\bar B^*B+c.c.)$ & $(3.5+7.7i)\%$ & $(3.0+8.5i)\%$ & $(3.2+9.4i)\%$ & $(3.0+9.7i)\%$ & $(0.5+7.7i)\%$ \\
\hline
$\mathcal P(\bar B^*B^*(^1P_1))$ & $(78.2-7.0i)\%$ & $(77.1-9.4i)\%$ & $(75.1-13.0i)\%$ & $(70.8-16.3i)\%$ & $(63.3-16.2i)\%$ \\
\hline
$\mathcal P(\bar B^*B^*(^5P_1))$ & $(18.3-0.7i)\%$ & $(19.9+0.9i)\%$ & $(21.7+3.6i)\%$ & $(26.2+6.6i)\%$ & $(36.2+8.5i)\%$ \\
\hline
$|\mathrm{Res}|(\bar B^*B+c.c.)$ (GeV$^{-1}$) & 0.208 & 0.212 & 0.226 & 0.223 & 0.168 \\
\hline
$|\mathrm{Res}|(\bar B^*B^*(^1P_1))$ (GeV$^{-1}$) & 0.555 & 0.511 & 0.520 & 0.488 & 0.338 \\
\hline
$|\mathrm{Res}|(\bar B^*B^*(^5P_1))$ (GeV$^{-1}$) & 0.269 & 0.260 & 0.287 & 0.310 & 0.264 \\
\bottomrule[0.8pt]
\end{tabular}
\end{table*}

The channel composition in Table~\ref{tab:10650} clarifies the dynamical origin of this pole. Since the complex-scaled Hamiltonian is non-Hermitian, the channel weights are evaluated using the c-product normalization of the wave function~\cite{Lin:2023ihj}, i.e.,
\begin{eqnarray}
(\phi|\phi)=
\sum_i\int\frac{d^3{\bm p}}{(2\pi)^3}
e^{-3i\theta}\phi_i(\tilde{\bm p})^2=1,
\end{eqnarray}
%where $\theta$ is the rotation angle in the complex scaling method and $i$ labels the coupled channels.
Over the full cutoff range, the $B^*\bar B^*(^1P_1)$ component is the largest one, while the $B^*\bar B^*(^5P_1)$ component remains sizable, although it becomes less important as the pole approaches threshold. The $B^*\bar B+c.c.(^3P_1)$ component is smaller but nonvanishing, reflecting coupled-channel mixing with the lower open-bottom channel. The near-threshold behavior of $Y(10650)$, together with its dominant $B^*\bar B^*$ composition, provides the dynamical basis for the open-bottom and hidden-bottom analyses below.

\iffalse
\footnote{For the non-Hermitian Hamiltonian, the wave function normalization is defined by the c-product~\cite{Lin:2023ihj},
\begin{eqnarray}
(\phi|\phi)=
\sum_i\int\frac{d^3{\bm p}}{(2\pi)^3}
e^{-3i\theta}\phi_i(\tilde{\bm p})^2=1,
\end{eqnarray}
where $\theta$ is the rotation angle in the complex scaling method and $i$ labels the coupled channels. We define $\mathcal P_i=(\phi_i|\phi_i)$, which roughly reflects the weight of the $i$-th channel.}
\fi

\textit{Experimental test in Belle~(II).---}Having established a threshold-pinned $P$-wave $B^*\bar B^*$ pole, we now turn to its signatures in exclusive open-bottom final states of $e^+e^-$ annihilation. The pole position alone is insufficient: one also needs the residues, which determine how strongly the pole is coupled to the physical $B^*\bar B^*$ and $B\bar B^*$ channels. These residues are not introduced as phenomenological parameters, but are extracted from the complex-scaled pole wave function.

In the complex scaling method, the resonance wave function is square integrable under the c-product normalization. For a resonance pole at $E_R$, its momentum-space wave function in channel $j$ is
\begin{eqnarray}
\phi_j(k)&=&
\frac{1}
{E_R-m_{\rm th}^{j}-k^2/(2\mu_j)}
\sum_m
\int\frac{d^3p}{(2\pi)^3}
e^{-3i\theta} \nonumber\\
&&\times V_{jm}(k,pe^{-i\theta})
\tilde\phi_m(p),
\label{eq:phianalytic}
\end{eqnarray}
where $\tilde\phi_m(p)$ is the complex-scaled wave function and $m$ runs over the coupled channels. The integration runs along the complex-scaled path required by the analytic continuation. If the external momentum $k$ is further continued to the complex plane, Eq.~(\ref{eq:phianalytic}) shows that $\phi_j(k)$ develops a singularity at $k=k_{R,j}$, defined by
\begin{eqnarray}
E_R=m_{\rm th}^{j}+\frac{k_{R,j}^{2}}{2\mu_j}.
\label{eq:kR}
\end{eqnarray}
This wave-function singularity and the scattering-amplitude pole therefore have the same physical origin: the resonance pole at $E_R$.

The residue of this pole is determined by the outgoing-wave component of the asymptotic behavior of the coordinate-space wave function~\cite{Lin:2023ihj}. For channel $j$,
\begin{eqnarray}
\psi_j(r)\!\!\xrightarrow{\scriptstyle r\to\infty}\!
\frac{i^{l}k_{R,j}}{2\pi}
\!\!\lim_{p\to k_{R,j}}\!\!
(p-k_{R,j})\phi_j(p)
\frac{e^{i(k_{R,j}r-l\pi/2)}}{r}.
\label{eq:asymp}
\end{eqnarray}
From Eqs.~\eqref{eq:phianalytic} and \eqref{eq:asymp}, the outgoing-wave coefficient is equivalently expressed as the pole residue of the diagonal $T$ matrix~\cite{Lin:2023ihj},
\begin{eqnarray}
{\rm Res}_j
\!=\!
\lim_{\scriptscriptstyle E\to E_R}(E-E_R)T_{jj}(E)
=
\langle k_{R,j}|\hat V|\phi_R)^2 .
\label{eq:residue}
\end{eqnarray}
The coupling strengths of $Y(10650)$ to $B^*\bar B^*$ and $B\bar B^*$ are therefore fixed by the pole wave function itself.

For the process $e^+e^-\to Y(10650)\to (B^{(*)}\bar B^{(*)})_j$, the pole contribution to the cross sections can be written as%~\cite{Hanhart:2010wh}
\begin{eqnarray}
\frac{d\sigma_j}{dE}
=
\frac{ {\cal B}
\mu_j g_j(\Lambda) k_j(E)/(4\pi^2)
}{
\left|
E-E_Y
-\sum_i
\frac{\mu_i}{8\pi^2}
g_i(\Lambda)
\sqrt{-2\mu_i(E-m_{\rm th}^i)}
\right|^2
},
\label{eq:xsec}
\end{eqnarray}
where $i$ runs over the coupled open-bottom channels, $j$ denotes the observed final state, $g_j(\Lambda)=|{\rm Res}_j|$ is the effective coupling constant, $k_j(E)$ is the corresponding on-shell momentum, and ${\cal B}$ absorbs the short-distance production strength of the pole in $e^+e^-$ annihilation. The coupled-channel pole calculation can determine $E_Y$ and the residues $g_i(\Lambda)$, but not the overall production factor ${\cal B}$.

\begin{figure}[h]
\centering
\includegraphics[width=0.43\textwidth]{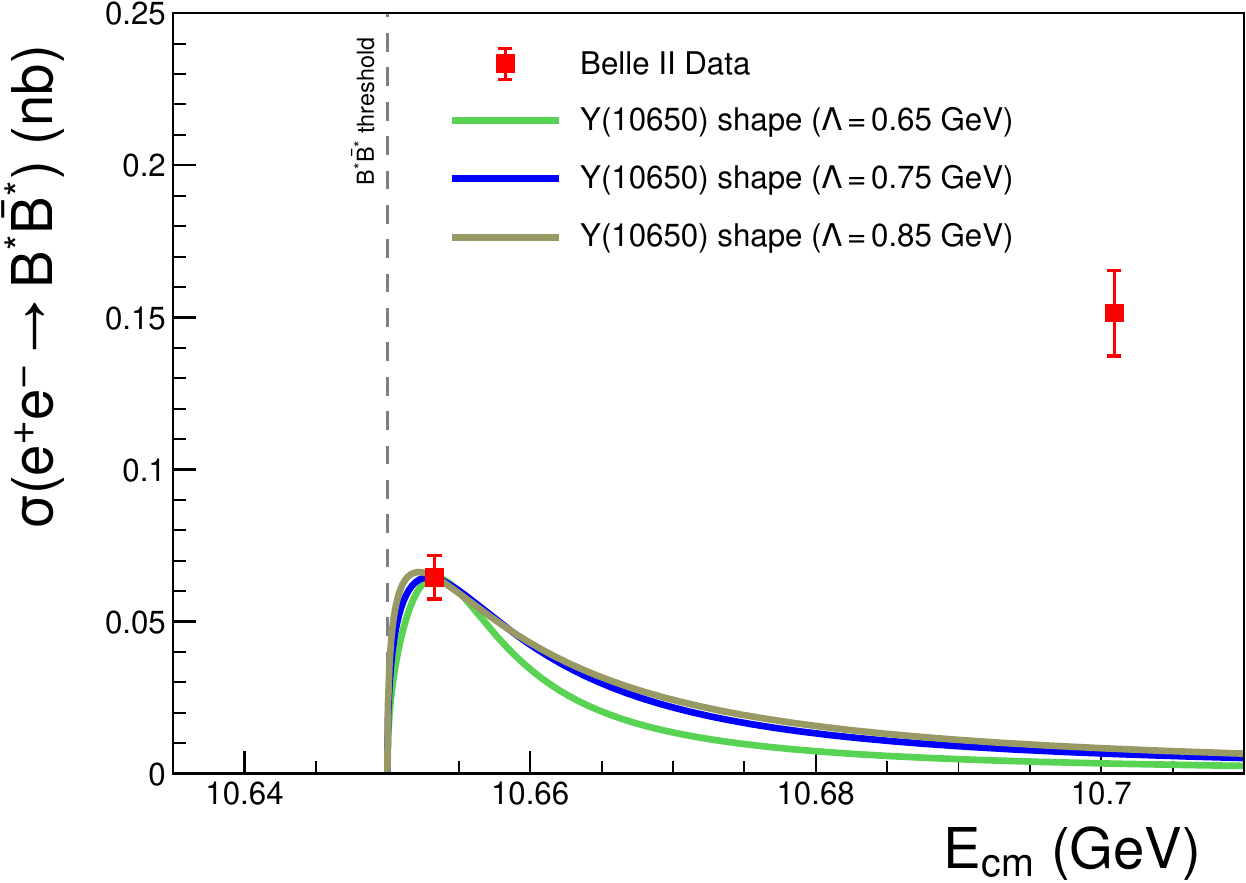}
\caption{
The description of the observed channel-opening enhancement for the cross sections of $e^+e^-\to B^*\bar B^*$~\cite{Belle-II:2024niz}.  %The red points are Belle II data~\cite{Belle-II:2024niz}, and the dashed line marks the $B^*\bar B^*$ threshold.  
The curves show the pole contributions for three representative $Y(10650)$ schemes.
}
\label{fig:bstarbstar_threshold}
\end{figure}

We then turn to a comparison with existing experimental data. Recent measurements by the Belle and Belle II Collaborations provide valuable information on the relevant open-bottom final states. The cross section distribution of $e^+e^-\to B\bar B^*$ shows a potential structure near the $B^*\bar B^*$ threshold. In $e^+e^-\to B^*\bar B^*$, Belle II observes a pronounced threshold enhancement, with the cross section already reaching the typical level of $0.1$ nb at only $3.7$ MeV above threshold~\cite{Belle:2021lzm,Belle-II:2024niz}. 
%Such a sharp enhancement near the opening of the $B^*\bar B^*$ threshold is difficult to reconcile with a smooth $P$-wave continuum contribution and points to a nearby pole contribution. 
As shown in Fig.~\ref{fig:bstarbstar_threshold}, the predicted $Y(10650)$ pole naturally accounts for this channel-opening enhancement.

The unknown production factor ${\cal B}$ is common to the open-bottom decay channels of the same pole.
For each pole scheme, the pole contribution in $e^+e^-\to B^*\bar B^*$ is specified by the pole position, the pole residue to $B^*\bar B^*$, and ${\cal B}$, with the first two fixed by the coupled-channel calculation listed in Table~\ref{tab:10650}. We determine ${\cal B}$ by matching the calculated pole contribution of $Y(10650)$ to the threshold-enhancement data point~\cite{Belle:2021lzm,Belle-II:2024niz}, located $3.7$ MeV above the $B^*\bar B^*$ threshold, as illustrated in Fig.~\ref{fig:bstarbstar_threshold}. This calibration is motivated by the strong suppression of the smooth $P$-wave open-bottom continuum near the threshold, where the phase-space factor is proportional to the third power of the final-state momentum {$(p^3)$}, while the observed enhancement appears immediately after the $B^*\bar B^*$ channel opens.
Once this common production factor is fixed, 
the absolute contribution to $e^+e^-\to Y(10650)\to B\bar B^*$ is predicted by Eq.~\eqref{eq:xsec} using the calculated pole residues to $B\bar B^*$ and corresponding phase-space factors.
In this sense, the $Y(10650)$ contribution in the $e^+e^- \to B\bar B^*$ is a correlated consequence of the same pole that accounts for the $B^*\bar B^*$ threshold enhancement, rather than an independently adjustable input.

\begin{figure}[!ht]
\begin{centering}
\scalebox{1.0}{\includegraphics[width=\linewidth]{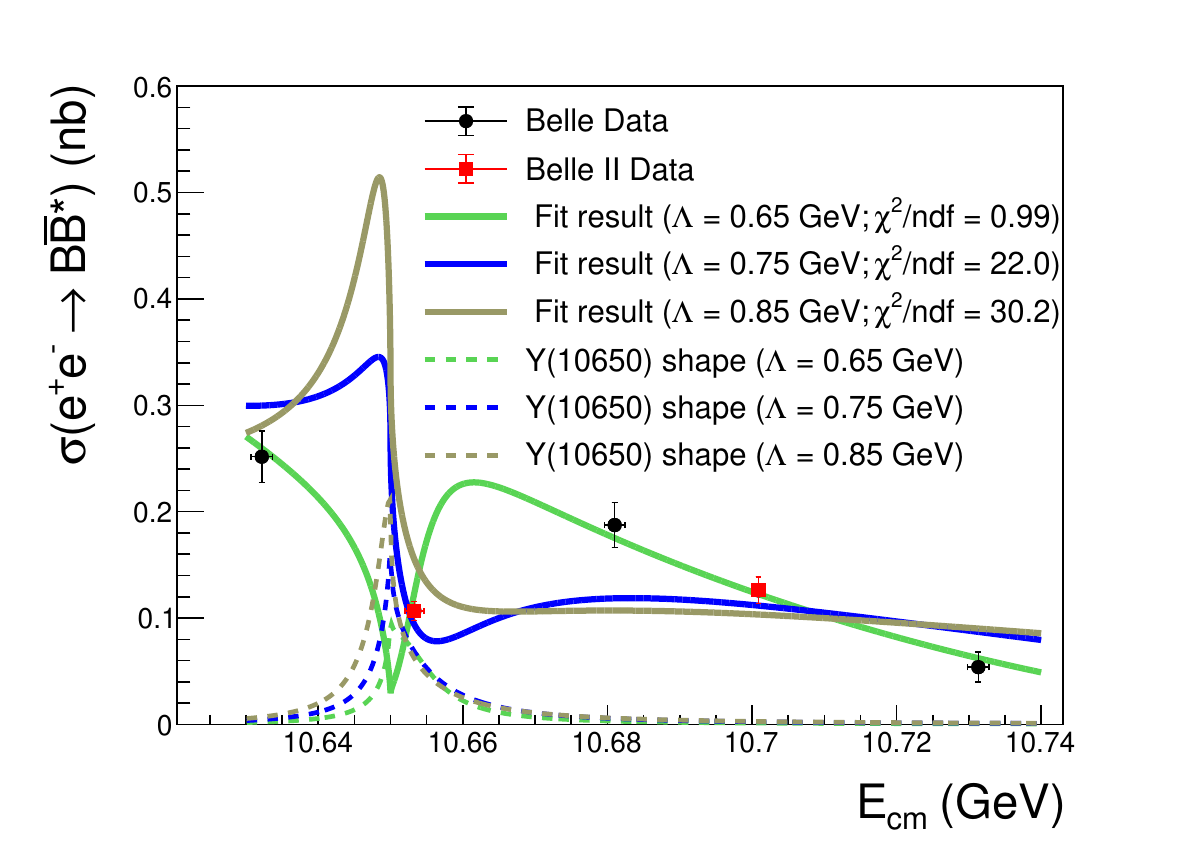}}
\caption{Absolute pole contributions to $e^+e^-\to Y(10650)\to B\bar B^*$ predicted from the $B^*\bar B^*$ threshold enhancement (dashed curves) and the corresponding fits (solid curves) to the measured $e^+e^-\to B\bar B^*$ cross sections~\cite{Belle:2021lzm,Belle-II:2024niz} for three representative pole schemes.}
\label{fig:Comparison}
\end{centering}
\end{figure}

Unlike the $B^*\bar B^*$ channel, the dip in the $\sigma(e^+e^-\to B\bar B^*)$ distribution around the $B^*\bar B^*$ threshold indicates interference between the $Y(10650)$ and continuum contributions~\cite{Belle:2021lzm,Belle-II:2024niz}. We therefore fit the $\sigma(e^+e^-\to B\bar B^*)$ distribution by including the $Y(10650)$ contribution, a smooth continuum background, and their interference. The shape and absolute magnitude of the $Y(10650)$ contribution are fixed by the procedure described above, while the continuum background is parametrized by a second-order Chebyshev polynomial. The coefficients of the Chebyshev polynomial and the interference angle are treated as free parameters in the fit. The fit results, together with the corresponding $\chi^2/\mathrm{ndf}$, are shown in Fig.~\ref{fig:Comparison}. We perform the fits with different cutoff schemes ($\Lambda=0.65,0.75,0.85$ GeV) and find that the $Y(10650)$ shape obtained with $\Lambda=0.65$ GeV describes the data well, with the %statistical 
significance of the $Y(10650)$ contribution reaching $6.7\sigma$. The fits with the $Y(10650)$ shapes obtained for other cutoff values are of poorer quality, although their corresponding significances remain above $3\sigma$. { Unlike the total fits, pure pole contributions (dashed curves) actually exhibit weak cutoff dependence. The $\Lambda=0.65$ GeV scheme performs better merely because it yields a smaller, better-matching overall magnitude.} The significances are estimated from the negative log-likelihood ratio $\sqrt{-2\ln({\cal L}_0/{\cal L}_{\rm max})}$. %with the change in degrees of freedom $\Delta{\rm ndf}=1$ associated with the interference phase between the $Y(10650)$ and continuum amplitudes. 
Here, ${\cal L}_0$ and ${\cal L}_{\rm max}$ are the maximized likelihoods of the fits without and with the $Y(10650)$ contribution, respectively.

This strong correlation between the predicted open-bottom line shapes and the observed enhancement or dip structures in the $\sigma(e^{+}e^{-} \to B^*\bar B^*)$ and $\sigma(e^{+}e^{-} \to B\bar B^*)$ distributions provides compelling support for the predicted $Y(10650)$. It also makes the scenario directly testable: future Belle II measurements with finer energy scans around the $B^*\bar B^*$ threshold can confirm the existence of $Y(10650)$ and verify the predicted pole pattern in this Letter.
%previous version"
%verify whether the correlated structures in the two open-bottom channels follow the predicted pole pattern.

\begin{figure}[t]
\begin{centering}
\scalebox{0.95}{\includegraphics[width=\linewidth]{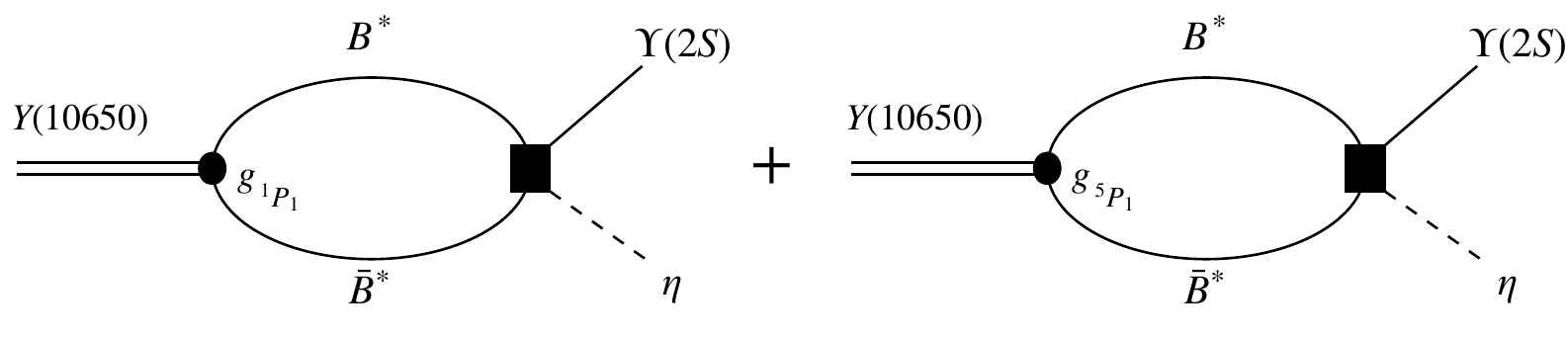}}
\caption{Schematic diagrams for $Y(10650)\to\Upsilon(2S)\eta$ through the near-threshold $B^*\bar B^*$ loop mechanism. The two diagrams correspond to the $YB^*\bar B^*$ source in the $^1P_1$ and $^5P_1$ configurations.}
\label{fig:decaydiag}
\end{centering}
\end{figure}

From another perspective, the pole identified above is dominated by the near-threshold $B^*\bar B^*$ dynamics and therefore coupled to hidden-bottom final states through bottom-meson loops. We consider $Y(10650)\to \Upsilon(2S)\eta$ via the $B^*\bar B^*$ loop mechanism shown in Fig.~\ref{fig:decaydiag}. Since $Y(10650)$ lies very close to the $B^*\bar B^*$ threshold, it is appropriate to treat the intermediate bottom mesons nonrelativistically. The $YB^*\bar B^*$ vertex is governed by the two $P$-wave configurations and the corresponding source tensor can be written as
\begin{eqnarray}
X^{\nu\rho}({\bm l})
&=&
i g_Y^{(1)}\delta^{\nu\rho}({\bm\epsilon}_Y\cdot {\bm l}) \nonumber\\
&+&i g_Y^{(5)}
\left[
\epsilon_Y^\nu l^\rho+
\epsilon_Y^\rho l^\nu
-\frac{2}{3}\delta^{\nu\rho}({\bm\epsilon}_Y\cdot {\bm l})
\right],
\label{eq:Ysource}
\end{eqnarray}
where ${\bm l}$ is the loop momentum, and $\nu$ and $\rho$ correspond to the vector meson $B^*$ and $\bar B^*$, respectively. The first and second terms describe the $B^*\bar B^*(^1P_1)$ and $B^*\bar B^*(^5P_1)$ sources, respectively, and the relevant coupling constants are connected to pole residues.  Because the momentum transfer in the transition $B^*\bar B^*\to \Upsilon(2S)\eta$ is limited by the near-threshold kinematics, the short-distance transition can be encoded in local four-point operators. We write the most general amplitude linear in the loop momentum and in the $\eta$ momentum as
\begin{eqnarray}
i{\cal M}_{4pt}=i\sum_{a=1}^{7}c_a{\cal O}_a,
\label{eq:4pt}
\end{eqnarray}
where the seven independent operators span the allowed spin-momentum structures. An example operator is $(\bm\epsilon_\Upsilon^*\cdot\bm\epsilon_1)\bm l\cdot(\bm\epsilon_2\times\bm p_\eta)$, with $\bm\epsilon_{1,2}$ as the polarizations of the intermediate $B^*$, $\bar B^*$. The remaining operators are given in the SM~\cite{suppl}. 
%In our numerical evaluation, 
The coefficients $c_a$ are fixed by matching this local amplitude to the $t$- and $u$-channel $B$ and $B^*$ meson exchange amplitudes. 
%In the present isoscalar channel, the symmetrized $t$- and $u$-channel kernels give identical contributions after contraction with the symmetric $YB^*\bar B^*$ source.

After matching and performing the loop integral, the total amplitude takes the form
\begin{eqnarray}
{\cal A}^{\rm tot}
=
{\cal N}^{\rm tot}
({\bm\epsilon}_Y\times{\bm\epsilon}_\Upsilon^*)\cdot{\bm p}_\eta,
\label{eq:Atotform}
\end{eqnarray}
with
\begin{eqnarray}
{\cal N}^{\rm tot}
&=&
-\frac{1}{2m_{B^*}^2}I_2(m_Y,\alpha_p)
\bigg[
\widehat C_{B^*}
\left(
\frac{4}{3}g_Y^{(1)}+
\frac{10}{9}g_Y^{(5)}
\right) \nonumber\\
&&+
\widehat C_{B}
\left(
\frac{2}{3}g_Y^{(1)}-
\frac{10}{9}g_Y^{(5)}
\right)
\bigg].
\label{eq:Atot}
\end{eqnarray}
Here $I_2(m_Y,\alpha_p)$ is the regulated loop integral, and $\widehat C_{B^*}$ and $\widehat C_{B}$ encode the effective $B^*$ and $B$ meson exchange coefficients for the $\Upsilon(2S)\eta$ channel, respectively,  whose definitions can be found in the SM~\cite{suppl}. 

{ Here, we take a sharp cutoff $\alpha_p=1.0$--$1.3~\mathrm{GeV}$ of the order of the typical hadronic scale~\cite{Wang:2020dko,Oset:2022xji,Baru:2016iwj,Wang:2023hpp}. This range also allows for the
nonrelativistic treatment of the intermediate bottom mesons.  For the three representative pole schemes above, with $E_{\mathrm{pole}}=10653.5-0.2i$ MeV, $10652.9-1.3i$ MeV and $10651.0-2.6i$ MeV, the predicted partial widths $\Gamma[Y(10650)\to\Upsilon(2S)\eta]$ are $12$--$54$ keV, $23$--$108$ keV and $25$--$117$ keV, respectively. These results place the expected width at the
$\mathcal{O}(10~\mathrm{keV})\sim \mathcal{O}(100~\mathrm{keV})$ scale. The estimate
should be viewed as a natural width scale rather than a precision prediction. }

% The hidden-bottom channel provides an independent test for this picture. Using the same pole schemes, we evaluate the cross sections of $e^+e^-\to Y(10650)\to\Upsilon(2S)\eta$ and {\color{red} find that the recently measured Belle II cross sections at the central value are reproduced for
% $\Gamma[Y(10650)\to\Upsilon(2S)\eta]\simeq 100~\mathrm{keV}$,
% with the width range extending from several tens to a few hundred keV when the
% experimental uncertainties are taken into account, as shown in
% Fig.~\ref{fig:u2eta}. }
{
The recently measured Belle II cross sections~\cite{Belle-II:2025ubm} are reproduced for $Y(10650)\to\Upsilon(2S)\eta$. 
The measured $\Gamma[Y(10650)\to\Upsilon(2S)\eta]$ has a central value around 100 keV, and ranges from several tens to a few hundred keV when the
experimental uncertainties are taken into account, as shown in
Fig.~\ref{fig:u2eta}. 
} 
This scale emerges { very} naturally from the loop calculation {and thus provides an independent test for this picture}. %The $Y(10650)$ therefore accounts for the correlated channel-opening enhancement in $\sigma(e^+e^- \to B^*\bar B^*)$, the accompanying structure in $\sigma(e^+e^- \to B\bar B^*)$ distribution, and the measured $\sigma(e^+e^- \to \Upsilon(2S)\eta)$ near the $B^*\bar B^*$ threshold. 
%These results provide a set of mutually connected open- and hidden-bottom signatures of $Y(10650)$, making it a compelling candidate for the first neutral isoscalar exotic bottomoniumlike state.

In summary, we predict a new $J^{PC}=1^{--}$ bottomoniumlike pole, $Y(10650)$, generated by $P$-wave $B^{(*)}\bar B^{(*)}$ coupled-channel dynamics. Its unusual RS location, proximity to the $B^*\bar B^*$ threshold, and dominant $B^*\bar B^*$ composition make it qualitatively different from a conventional bottomonium excitation. The same state gives a correlated account of the sharp threshold enhancement in $\sigma(e^+e^- \to B^*\bar B^*)$, the structure close to the $B^*\bar B^*$ threshold in $\sigma(e^+e^- \to B\bar B^*)$ distribution, and the recently measured $e^+e^- \to \Upsilon(2S)\eta$ cross sections. These results support $Y(10650)$ as a compelling candidate for the first neutral isoscalar exotic hadron in the bottomonium sector and demonstrate that higher partial wave near-threshold dynamics can generate experimentally accessible exotic hadrons. Finer Belle II scans around the $B^*\bar B^*$ threshold, together with searches for correlated structures in additional open- and hidden-bottom final states, will provide decisive tests for this mechanism. 

\section*{Acknowledgement}

This project was supported by the National Natural Science Foundation of China under Grants No. 11975033, No. 12147168, No. 12070131001, No. 12105072, No. 12475137, No. 12405088, No. 12547101, No. 12405160, No. 12135005, and No. 12405102, and the National Key Research and Development Program under the contract No. 2024YFA1610503. This project was also funded by the Deutsche Forschungsgemeinschaft (DFG,
German Research Foundation, Project ID 196253076-TRR 110), Start-up Funds of Southeast University (Grant No. 4007022506) and the Natural Science Foundation of Shandong Province (Grants No. ZR2024QA041). J. Z. Wang is also supported by the Start-up Funds of Chongqing University.

\bibliography{refs}

% =========================
% End Matter
% =========================

\clearpage

\onecolumngrid
\vspace*{-1.5em}
\begin{center}
{\bfseries\large End Matter}
\end{center}
\vspace{1.5em}
\twocolumngrid

\appendix

\iffalse
\textit{Appendix: $Y(10650)$ pole interpretation of channel-opening enhancement in $e^+e^-\to B^*\bar B^*$.---}Figure~\ref{fig:bstarbstar_threshold} shows how the predicted $Y(10650)$ pole
accounts for the enhancement phenomenon when the $B^*\bar B^*$ channel opens in
$e^+e^-$ annihilation.  For each pole scheme, the pole contribution is
determined by the pole position, the $B^*\bar B^*$ residues and production strength, where the first two can be fixed by the
coupled-channel calculation. The remaining short-distance production strength in $e^+e^-$ annihilation is fixed by matching the calculated pole contribution to the Belle II enhancement closest to the $B^*\bar B^*$ threshold. This calibration is justified only in the immediate threshold region, where the nonresonant $P$-wave continuum is phase-space suppressed and should vary smoothly.  

It can be seen that all three representative pole schemes reproduce the observed channel-opening enhancement. Away from the threshold, smooth continuum production and higher vector bottomonium contributions are expected to become increasingly relevant. More importantly, the $B^*\bar B^*$ data determine the production strength of the $Y(10650)$ pole. This strength is then carried over, without further adjustment, to the analysis of $e^+e^-\to B\bar B^*$ within the same pole scheme in the main text.  The resulting structure in $B\bar B^*$ channel is
therefore a correlated prediction of the $B^*\bar B^*$ threshold enhancement rather than
an independently normalized signal.
\fi

\textit{Appendix A: The cross sections of $e^+e^-$ annihilation to $\Upsilon(2S)\eta$.---}The $Y(10650)$ contribution to $e^+e^-\to \Upsilon(2S)\eta$ can be described by the same form as $d\sigma_j/dE$ in Eq.~(\ref{eq:xsec}), with the pole residue and the phase-space factor in the open-bottom channel replaced by their counterparts in the $\Upsilon(2S)\eta$ channel. This hidden-bottom residue is fixed by the partial decay width $\Gamma[Y(10650)\to\Upsilon(2S)\eta]$. Thus, the chosen partial width sets the overall magnitude of the hidden-bottom cross sections of $e^+e^-\to Y(10650) \to \Upsilon(2S)\eta$, while the pole schemes of $Y(10650)$ only determine its line shape.

\begin{figure}[th]
\begin{centering}
\scalebox{1.35}{\includegraphics[width=0.72\linewidth]{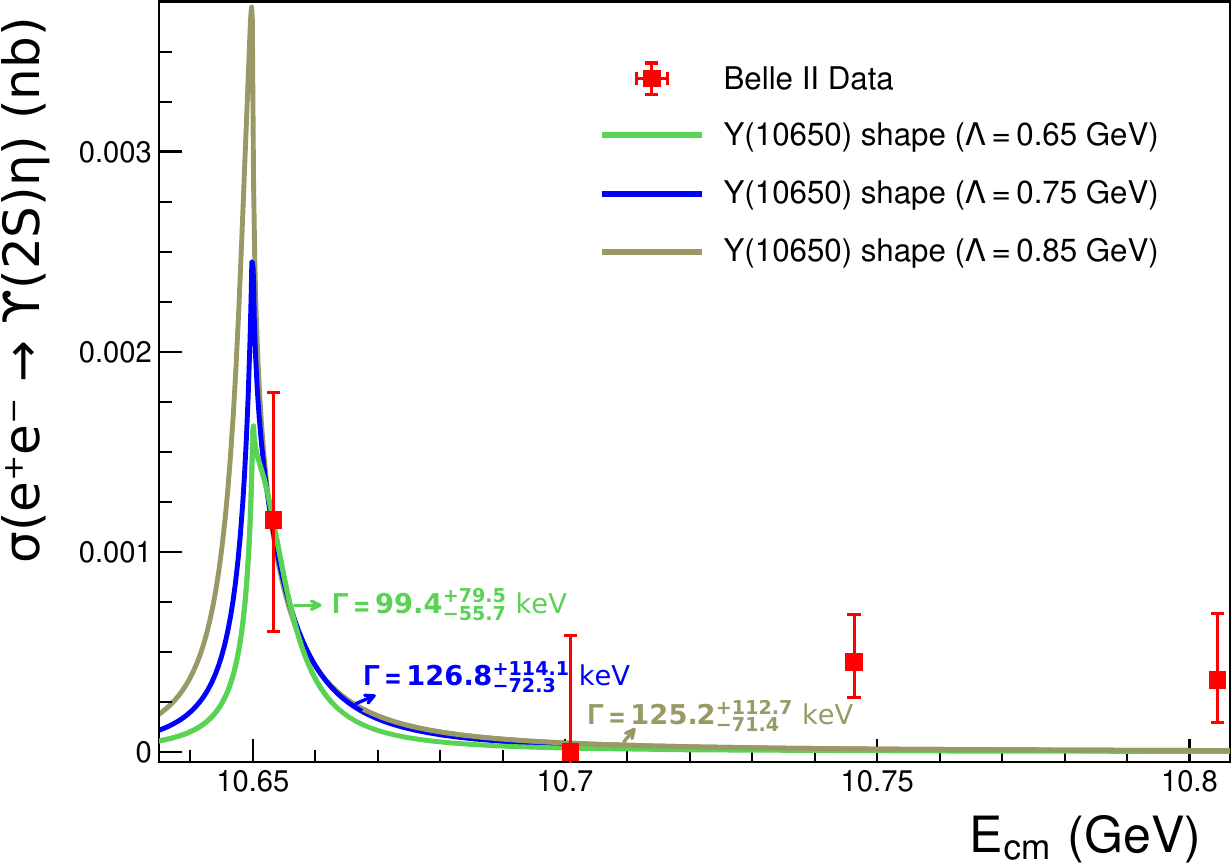}}
\caption{Comparison between the predicted cross sections of $e^+e^-\to Y(10650)\to \Upsilon(2S)\eta$ and the Belle II data~\cite{Belle-II:2025ubm}. { Here, only the curves corresponding to the
central values of partial decay width $\Gamma[Y(10650)\to\Upsilon(2S)\eta$ are shown, whose uncertainties are obtained by propagating the experimental uncertainty of
the first data point. }  }
\label{fig:u2eta}
\end{centering}
\end{figure}

Figure~\ref{fig:u2eta} compares the resulting cross sections of $e^+e^-\to Y(10650)\to \Upsilon(2S)\eta$ with the Belle II data~\cite{Belle-II:2025ubm}. The three curves use the same pole schemes as those in the open-bottom analysis. {The assigned partial decay widths for the three pole schemes with 
$\Lambda=0.65, 0.75,$ and $0.85$ GeV,
$\Gamma[Y(10650)\to\Upsilon(2S)\eta]
=\,99.4^{+79.5}_{-55.7},\quad
126.8^{+114.1}_{-72.3},\quad
125.2^{+112.7}_{-71.4}\ {\rm keV}$,
respectively, reproduce the enhanced Belle II data point near the 
$B^*\bar B^*$ threshold within its experimental uncertainty. The uncertainties
of these partial widths are obtained by propagating the experimental uncertainty of
this data point. In Fig.~\ref{fig:u2eta}, only the curves corresponding to the
central width values are shown, since including the associated error bands would
obscure the comparison among the three pole schemes. These values are representative $\mathcal{O}(100~{\rm keV})$ widths, as expected from
the theoretical calculations of the $B^*\bar B^*$ loop mechanism.}  Since the absolute cross sections scale linearly with the assumed partial width, future precise measurements can directly constrain $\Gamma[Y(10650)\to\Upsilon(2S)\eta]$.

\newpage

%% attached the SM as the appendix

 \clearpage

 \setcounter{equation}{0}
\setcounter{figure}{0}
\setcounter{table}{0}
\setcounter{page}{1}
\makeatletter
\renewcommand{\theequation}{S\arabic{equation}}
\renewcommand{\thefigure}{S\arabic{figure}}
\renewcommand\thetable{SM-\Roman{table}}  

% \onecolumn
%\begin{widetext}

%\begin{center}
%\textbf{\large Supplemental Materials}
%\end{center}

\noindent\makebox[\textwidth][c]{\textbf{\large Supplemental Materials}}
\vspace{1em}

\iffalse
\begin{mdframed}[hidealllines=true,innerleftmargin=0.1\textwidth,innerrightmargin=0.1\textwidth]
~~~~~This supplemental material provides additional details on \smabs.
\end{mdframed}
\fi

%\end{widetext}

\onecolumngrid

This Supplemental Material provides the technical ingredients used in the main text.  Section~I gives the complete one-boson-exchange (OBE) interactions of the coupled-channel framework in the open-bottom sector, including the flavor factors entering the isoscalar  coupled-channel problem.  Section~II derives the local four-point amplitude of the hidden-bottom transition $B^*\bar B^*\to\Upsilon(2S)\eta$ and shows explicitly how the low-energy constants are fixed by the $B^*$ and $B$ meson exchange saturation used for the numerical estimates. Section~III summarizes the definitions of loop integrals and the decay width formula associated with  $Y(10650)\to\Upsilon(2S)\eta$. %Section~III summarizes the loop integral, the partial-width formula, and the connection between the calculated hidden-bottom width and the $e^+e^-\to\Upsilon(2S)\eta$ cross section.

\section{OBE interactions in the open-bottom sector}

The coupled-channel potentials of $B^{(*)}\bar B^{(*)} \to B^{(*)}\bar B^{(*)}$ are constructed from heavy-meson effective Lagrangians constrained by heavy-quark spin symmetry, chiral symmetry, and SU(2) flavor symmetry~\cite{Wise:1992hn,Yan:1992gz,Grinstein:1992qt,Casalbuoni:1996pg,Li:2012cs,Li:2012ss}.  We denote the bottom pseudoscalar and vector mesons by ${\cal B}$ and ${\cal B}^*$, respectively.  The exchanged light mesons are the pseudoscalars $\pi$ and $\eta$, the scalar $\sigma$, and the vectors $\rho$ and $\omega$.  The effective Lagrangians are
\begin{eqnarray}
{\cal L}_{{\cal B}^{(*)}{\cal B}^{(*)}\sigma}
&=&-2g_s {\cal B}_b^\dag {\cal B}_b\sigma
+2g_s {\cal B}_b^*\cdot {\cal B}_b^{*\dag}\sigma,
\nonumber\\
{\cal L}_{{\cal B}^{(*)}{\cal B}^{(*)}P}
&=&\frac{2g}{f_\pi}
\left({\cal B}_b {\cal B}_{a\lambda}^{*\dag}
+{\cal B}_{b\lambda}^{*}{\cal B}_a^\dag\right)\partial^\lambda P_{ba}
+i\frac{2g}{f_\pi}v^\alpha\varepsilon_{\alpha\mu\nu\lambda}
{\cal B}_b^{*\mu}{\cal B}_a^{*\lambda\dag}\partial^\nu P_{ba},
\nonumber\\
{\cal L}_{{\cal B}^{(*)}{\cal B}^{(*)}V}
&=&\sqrt2\beta g_V {\cal B}_b {\cal B}_a^\dag v\cdot V_{ba}
-2\sqrt2\lambda g_V\epsilon_{\lambda\mu\alpha\beta}v^\lambda
\left({\cal B}_b{\cal B}_a^{*\mu\dag}+{\cal B}_b^{*\mu}{\cal B}_a^\dag\right)
\partial^\alpha V_{ba}^\beta
\nonumber\\
&&-\sqrt2\beta g_V {\cal B}_b^*\cdot {\cal B}_a^{*\dag}v\cdot V_{ba}
-i2\sqrt2\lambda g_V {\cal B}_b^{*\mu}{\cal B}_a^{*\nu\dag}
(\partial_\mu V_\nu-\partial_\nu V_\mu)_{ba}.
\label{eq:SMlag}
\end{eqnarray}
The numerical couplings are taken as $g=0.46$~\cite{Workman:2022ynf,Khodjamirian:2020mlb}, $f_\pi=132$ MeV, $g_V=m_\rho/f_\pi=5.8$~\cite{Isola:2003fh,Bando:1987br}, $\lambda=0.16~{\rm GeV}^{-1}$, $\beta=0.9$~\cite{Isola:2003fh,Bando:1987br}, and $g_s=g_{\sigma NN}/3=4.06$~\cite{Riska:2000gd,Durso:1980vn,Wu:2023uva}.  In SU(2) flavor space we use
\begin{eqnarray}
P=\begin{pmatrix}
{\pi^0}/{\sqrt2}+{\eta}/{\sqrt6} & \pi^+\\
\pi^- & -{\pi^0}/{\sqrt2}+{\eta}/{\sqrt6}
\end{pmatrix},\qquad
V=\begin{pmatrix}
{\rho^0}/{\sqrt2}+{\omega}/{\sqrt2} & \rho^+\\
\rho^- & -{\rho^0}/{\sqrt2}+{\omega}/{\sqrt2}
\end{pmatrix}.
\label{eq:SMmatrices}
\end{eqnarray}

The effective potentials or scattering kernels in momentum space are obtained in the Breit approximation.  We define ${\bm q}={\bm p}_1-{\bm p}_3$ and ${\bm k}={\bm p}_1+{\bm p}_3$, where ${\bm p}_1, {\bm p}_2$ and ${\bm p}_3, {\bm p}_4$ are the initial and final heavy-meson momenta.  The polarization vectors of the external vector mesons are denoted by ${\bm\epsilon}_i$.  The nonlocal monopole regulator is used in the main text,
\begin{eqnarray}
{\cal F}(p^2,p^{\prime 2})=
\frac{\Lambda^2}{\Lambda^2+p^2}
\frac{\Lambda^2}{\Lambda^2+p^{\prime 2}},
\label{eq:SMreg}
\end{eqnarray}
which is multiplied after projecting the interaction kernels to the relevant partial waves.  We list the explicit unprojected interaction potentials below.
For $B(p_1)\bar B(p_2)\to B(p_3)\bar B(p_4)$ and $B(p_1)\bar B(p_2)\to B^*(p_3)\bar B^*(p_4)$,
\begin{eqnarray}
{\cal V}^{B\bar B\to B\bar B}_{\sigma}
&=&-g_s^2\frac{1}{{\bm q}^2+m_\sigma^2}C_\sigma,
\nonumber\\
{\cal V}^{B\bar B\to B\bar B}_{\rho/\omega}
&=&-\frac{\beta^2g_V^2}{2}\frac{1}{{\bm q}^2+m_{\rho/\omega}^2}C_{\rho/\omega},
\nonumber\\
{\cal V}^{B\bar B\to B^*\bar B^*}_{\pi/\eta}
&=&\frac{g^2}{f_\pi^2}
\frac{({\bm\epsilon}_3^\dag\cdot{\bm q})({\bm\epsilon}_4^\dag\cdot{\bm q})}
{{\bm q}^2+m_{\pi/\eta}^2}C_{\pi/\eta},
\nonumber\\
{\cal V}^{B\bar B\to B^*\bar B^*}_{\rho/\omega}
&=&2\lambda^2g_V^2
\frac{({\bm\epsilon}_3^\dag\times{\bm q})\cdot({\bm\epsilon}_4^\dag\times{\bm q})}
{{\bm q}^2+m_{\rho/\omega}^2}C_{\rho/\omega}.
\label{eq:SMVBB}
\end{eqnarray}
For the $C$-parity eigenstate $[B\bar B^*]$, which simultaneously involves the component of $B\bar B^*$ and $\bar BB^*$, the effective potentials of $[B(p_1)\bar B^*(p_2)]\to [B(p_3)\bar B^*(p_4)]$ are
\begin{eqnarray}
V_\sigma^D
&=&-\frac{g_s^2}{{\bm q}^2+m_\sigma^2}C_\sigma^D,
\nonumber\\
V_{\pi/\eta}^C
&=&-\frac{g^2}{f_\pi^2}
\frac{({\bm\epsilon}_2\cdot{\bm k})({\bm\epsilon}_4^\dag\cdot{\bm k})}
{{\bm k}^2-k_0^2+m_{\pi/\eta}^2}C_{\pi/\eta}^C,
\nonumber\\
V_{\rho/\omega}^D
&=&\frac{1}{2}\beta^2g_V^2
\frac{{\bm\epsilon}_2\cdot{\bm\epsilon}_4^\dag}
{{\bm q}^2+m_{\rho/\omega}^2}C_{\rho/\omega}^D,
\nonumber\\
V_{\rho/\omega}^C
&=&\lambda^2g_V^2
\frac{({\bm\epsilon}_2\cdot{\bm k})({\bm\epsilon}_4^\dag\cdot{\bm k})
-{\bm k}^2({\bm\epsilon}_2\cdot{\bm\epsilon}_4^\dag)}
{{\bm k}^2-k_0^2+m_{\rho/\omega}^2}C_{\rho/\omega}^C.
\label{eq:SMVBBstar}
\end{eqnarray}
The effective potentials for $[B(p_1)\bar B^*(p_2)]\to B^*(p_3)\bar B^*(p_4)$ are
\begin{eqnarray}
{\cal V}^{[B\bar B^*]\to B^*\bar B^*}_{\pi/\eta}
&=&\frac{i g^2}{f_\pi^2}
\frac{({\bm\epsilon}_3^\dag\cdot{\bm q})
[({\bm\epsilon}_4^\dag\times{\bm q})\cdot{\bm\epsilon}_2]}
{{\bm q}^2+m_{\pi/\eta}^2}C'_{\pi/\eta},
\nonumber\\
{\cal V}^{[B\bar B^*]\to B^*\bar B^*}_{\rho/\omega}
&=&2i\lambda^2g_V^2
\left [\frac{({\bm\epsilon}_3^\dag\times{\bm q})\cdot{\bm\epsilon}_4^\dag
({\bm\epsilon}_2\cdot{\bm q})}
{{\bm q}^2+m_{\rho/\omega}^2}-\frac{({\bm\epsilon}_3^\dag\times{\bm q})\cdot{\bm\epsilon}_2
({\bm\epsilon}_4^\dag\cdot{\bm q})}
{{\bm q}^2+m_{\rho/\omega}^2}\right ]C'_{\rho/\omega}.
\label{eq:SMVtrans}
\end{eqnarray}
Finally, the effective potentials of $B^*(p_1)\bar B^*(p_2)\to B^*(p_3)\bar B^*(p_4)$ read
\begin{eqnarray}
{\cal V}^{B^*\bar B^*\to B^*\bar B^*}_{\sigma}
&=&-g_s^2
\frac{({\bm\epsilon}_1\cdot{\bm\epsilon}_3^\dag)
({\bm\epsilon}_2\cdot{\bm\epsilon}_4^\dag)}{{\bm q}^2+m_\sigma^2}C_\sigma,
\nonumber\\
{\cal V}^{B^*\bar B^*\to B^*\bar B^*}_{\pi/\eta}
&=&\frac{g^2}{f_\pi^2}
\frac{[({\bm\epsilon}_1\times{\bm\epsilon}_3^\dag)\cdot{\bm q}]
[({\bm\epsilon}_2\times{\bm\epsilon}_4^\dag)\cdot{\bm q}]}
{{\bm q}^2+m_{\pi/\eta}^2}C_{\pi/\eta},
\nonumber\\
{\cal V}^{B^*\bar B^*\to B^*\bar B^*}_{\rho/\omega}
&=&\left\{
\frac{\beta^2g_V^2}{2}
\frac{({\bm\epsilon}_1\cdot{\bm\epsilon}_3^\dag)({\bm\epsilon}_2\cdot{\bm\epsilon}_4^\dag)}
{{\bm q}^2+m_{\rho/\omega}^2}
+\frac{2\lambda^2g_V^2}{{\bm q}^2+m_{\rho/\omega}^2}
\Big[
({\bm\epsilon}_1\cdot{\bm\epsilon}_2)({\bm\epsilon}_3^\dag\cdot{\bm q})({\bm\epsilon}_4^\dag\cdot{\bm q})
\right.
\nonumber\\
&&\left.
-({\bm\epsilon}_1\cdot{\bm\epsilon}_4^\dag)({\bm\epsilon}_3^\dag\cdot{\bm q})({\bm\epsilon}_2\cdot{\bm q})
-({\bm\epsilon}_2\cdot{\bm\epsilon}_3^\dag)({\bm\epsilon}_4^\dag\cdot{\bm q})({\bm\epsilon}_1\cdot{\bm q})
\right.
\nonumber\\
&&\left.
+({\bm\epsilon}_3^\dag\cdot{\bm\epsilon}_4^\dag)({\bm\epsilon}_1\cdot{\bm q})({\bm\epsilon}_2\cdot{\bm q})
\Big]\right\}C_{\rho/\omega}(-1).
\label{eq:SMVBstarstar}
\end{eqnarray}
The isoscalar flavor factors entering the above interaction kernels are listed in Table~\ref{tab:flavor}.  Different superscripts are retained because they refer to different potential elements and different transition structures: $D$ and $C$ denote a direct and crossed diagram in the $[B\bar B^*]$ channel, respectively. For the bottomed meson, the isospin breaking effect is tiny and the isospin averaged masses of the bottomed meson from the Particle Data Group (PDG) are taken~\cite{Workman:2022ynf}.

\begin{table}[t]
\caption{Isoscalar flavor factors entering the OBE potentials. Each row gives the coefficient appearing explicitly in Eqs.~\eqref{eq:SMVBB}--\eqref{eq:SMVBstarstar}.}
\label{tab:flavor}
\centering
\renewcommand{\arraystretch}{1.28}
\setlength{\tabcolsep}{38pt}
\begin{tabular}{lll}
\hline\hline
Potential element & Coefficient & Value \\
\hline
\multirow{5}{*}{$B\bar B\to B\bar B$, $B^*\bar B^*\to B^*\bar B^*$ and $B\bar B\to B^*\bar B^*$}
& $C_\sigma$ & $1$ \\
& $C_\pi$ & $3/2$ \\
& $C_\eta$ & $1/6$ \\
& $C_\rho$ & $3/2$ \\
& $C_\omega$ & $1/2$ \\
\hline
\multirow{7}{*}{$[B\bar B^*]_{C=-}\to [B\bar B^*]_{C=-}$}
& $C_\sigma^D$ & $1$ \\
& $C_\rho^D$ & $-3/2$ \\
& $C_\omega^D$ & $-1/2$ \\
& $C_\pi^C$ & $-3/2$ \\
& $C_\eta^C$ & $-1/6$ \\
& $C_\rho^C$ & $3/2$ \\
& $C_\omega^C$ & $1/2$ \\
\hline
\multirow{4}{*}{$[B\bar B^*]_{C=-}\to B^*\bar B^*$}
& $C'_\pi$ & $3/\sqrt2$ \\
& $C'_\eta$ & $1/(3\sqrt2)$ \\
& $C'_\rho$ & $3/\sqrt2$ \\
& $C'_\omega$ & $1/\sqrt2$ \\
\hline\hline
\end{tabular}
\renewcommand{\arraystretch}{1.0}
\setlength{\tabcolsep}{6pt}
\end{table}

After the partial-wave projection of the OBE potentials listed in Eqs.~\eqref{eq:SMVBBstar}--\eqref{eq:SMVBstarstar}, the projected potentials in the $P$-wave basis with quantum number $J^{PC}=1^{--}$ can be obtained.  These projected potentials are then employed in the complex-scaled coupled-channel Schr\"odinger equation discussed in the main text.

\section{Effective four-point amplitude for $B^*\bar B^*\to\Upsilon(2S)\eta$}

The hidden-bottom transition of $Y(10650)$ can provide a probe of the near-threshold $B^*\bar B^*$ dynamics encoded in the predicted pole, because it proceeds through rescattering of the intermediate $B^*\bar B^*$ component into hidden-bottom final states. Since the pole lies only a few MeV from the $B^*\bar B^*$ threshold, the loop momentum carried by the intermediate bottom mesons is nonrelativistic. The transition $B^*\bar B^*\to\Upsilon(2S)\eta$ can therefore be organized into a local four-point amplitude constructed from the loop momentum ${\bm l}$ and the external momentum ${\bm p}_\eta$ of $\eta$ meson. 
This local parametrization actually describes the short-distance transition of the $B^*\bar B^*$ rescattering into the hidden-bottom final state.

We consider
\begin{eqnarray}
B^*(p_1,\epsilon_1)+\bar B^*(p_2,\epsilon_2)
\to \Upsilon(2S)(p_3,\epsilon_3)+\eta(p_4),
\label{eq:SMproc}
\end{eqnarray}
with
\begin{eqnarray}
p_1=(E_1,{\bm l}),\qquad
p_2=(E_2,-{\bm l}),\qquad
p_3=(E_3,-{\bm p}_4),\qquad
p_4=(E_4,{\bm p}_4).
\end{eqnarray}
The most general local four-point amplitude at the order needed here is
\begin{eqnarray}
i{\cal M}_{4pt}=i\sum_{a=1}^7 c_a{\cal O}_a,
\label{eq:SM4pt}
\end{eqnarray}
where the seven independent operators are
\begin{eqnarray}
{\cal O}_1&=&({\bm\epsilon}_3^*\cdot{\bm\epsilon}_1)
{\bm l}\cdot({\bm\epsilon}_2\times{\bm p}_4),
\nonumber\\
{\cal O}_2&=&({\bm\epsilon}_3^*\cdot{\bm\epsilon}_2)
{\bm l}\cdot({\bm\epsilon}_1\times{\bm p}_4),
\nonumber\\
{\cal O}_3&=&({\bm\epsilon}_3^*\cdot{\bm l})
({\bm\epsilon}_1\times{\bm\epsilon}_2)\cdot{\bm p}_4,
\nonumber\\
{\cal O}_4&=&({\bm\epsilon}_1\cdot{\bm l})
({\bm\epsilon}_3^*\times{\bm\epsilon}_2)\cdot{\bm p}_4,
\nonumber\\
{\cal O}_5&=&({\bm\epsilon}_2\cdot{\bm l})
({\bm\epsilon}_3^*\times{\bm\epsilon}_1)\cdot{\bm p}_4,
\nonumber\\
{\cal O}_6&=&({\bm\epsilon}_1\cdot{\bm p}_4)
({\bm\epsilon}_3^*\times{\bm\epsilon}_2)\cdot{\bm l},
\nonumber\\
{\cal O}_7&=&({\bm\epsilon}_2\cdot{\bm p}_4)
({\bm\epsilon}_3^*\times{\bm\epsilon}_1)\cdot{\bm l}.
\label{eq:SMops}
\end{eqnarray}
In the present numerical analysis, the low-energy constants $c_a$ of the local four-point operators are fixed by matching to the $B$ and $B^*$ meson exchange amplitudes for $B^*\bar B^*\to\Upsilon(2S)\eta$. This matching provides a practical realization of the short-distance transition, without assuming it to be the unique microscopic mechanism.

\subsection{Matching to $B^*$-exchange amplitude}

We first consider the representative $t$-channel $B^*$-exchange contribution to
$B^*\bar B^*\to\Upsilon(2S)\eta$.  The exchanged momentum is
$q_t=p_1-p_3=p_4-p_2$, where $p_1$, $p_2$, $p_3$, and $p_4$ denote the
momenta of the incoming $B^*$, incoming $\bar B^*$, outgoing $\Upsilon(2S)$,
and outgoing $\eta$, respectively.

The $\Upsilon(2S)B^*\bar B^*$ vertex is written as~\cite{Achasov:1994vh,Deandrea:2003pv,Colangelo:2003sa,Lin:1999ad}
\begin{eqnarray}
{\cal V}_{\Upsilon B^*\bar B^*,t}^{\mu\nu\alpha}
=
i g_{\Upsilon B^*\bar B^*}
T_t^{\mu\nu\alpha},
\end{eqnarray}
with
\begin{eqnarray}
T_t^{\mu\nu\alpha}
=
-\,g^{\nu\alpha}(p_1^\mu-q_t^\mu)
-\,g^{\mu\alpha}q_t^\nu
+\,g^{\mu\nu}p_1^\alpha .
\label{eq:SMTtBstar}
\end{eqnarray}
Here $\mu$, $\nu$, and $\alpha$ label the outgoing $\Upsilon(2S)$, the
incoming $B^*$, and the exchanged $B^*$, respectively.  The coupling is
$g_{\Upsilon B^*\bar B^*}
=\frac{m_{\Upsilon(2S)}m_{B^*}}{f_{\Upsilon(2S)}m_B}$~\cite{Achasov:1994vh,Deandrea:2003pv,Colangelo:2003sa,Lin:1999ad} with $f_{\Upsilon(2S)}=0.497$ GeV~\cite{Huang:2018pmk}.  The $B^*\bar B^*\eta$ vertex is~\cite{Wise:1992hn,Yan:1992gz,Grinstein:1992qt,Casalbuoni:1996pg,Li:2012cs,Li:2012ss}
\begin{eqnarray}
{\cal V}_{B^*\bar B^*\eta,t}^{\rho\beta}
=
-i g_{B^*\bar B^*\eta}C_\eta\,
\epsilon_{\sigma\rho\beta\lambda}v^\sigma p_4^\lambda ,
\label{eq:SMetaBstarVertex4D}
\end{eqnarray}
where $\rho$ labels the incoming $\bar B^*$, $\beta$ labels the exchanged
$B^*$, $v^\sigma=(1,\bm 0)$, and
$g_{B^*\bar B^*\eta}=2g m_{B^*}/f_\pi$.  The heavy-meson velocity $v^{\sigma}$ selects
the spatial Levi-Civita tensor, giving
\begin{eqnarray}
{\cal V}_{B^*\bar B^*\eta,t}^{kb}
=
-i g_{B^*\bar B^*\eta}C_\eta\,
\epsilon^{kbm}p_4^m ,
\label{eq:SMetaBstarVertex3D}
\end{eqnarray}
where $k$ and $b$ are spatial indices.

The full $t$-channel exchange amplitude is
\begin{eqnarray}
i{\cal M}_{B^*,t}
=
\epsilon_{3\mu}^{*}\epsilon_{1\nu}\epsilon_{2\rho}
\,{\cal V}_{\Upsilon B^*\bar B^*,t}^{\mu\nu\alpha}
G_{\alpha\beta}(q_t)
{\cal V}_{B^*\bar B^*\eta,t}^{\rho\beta},
\label{eq:SMBstarFullAmp}
\end{eqnarray}
with
\begin{eqnarray}
G_{\alpha\beta}(q_t)
=
\frac{
i\left(-g_{\alpha\beta}
+q_{t\alpha}q_{t\beta}/m_{B^*}^2\right)
}{
q_t^2-m_{B^*}^2+i\epsilon
}.
\label{eq:SMBstarProp}
\end{eqnarray}
The longitudinal term $q_{t\alpha}q_{t\beta}/m_{B^*}^2$ gives the suppressed corrections of $1/m_{B^*}^2$ to the local amplitude and is neglected in
the leading matching.  Keeping the $-g_{\alpha\beta}$ part, one may separate
the contraction into its time and spatial components,
\begin{eqnarray}
-g_{\alpha\beta}
{\cal V}_{\Upsilon B^*\bar B^*,t}^{\mu\nu\alpha}
{\cal V}_{B^*\bar B^*\eta,t}^{\rho\beta}
=
-{\cal V}_{\Upsilon B^*\bar B^*,t}^{\mu\nu 0}
{\cal V}_{B^*\bar B^*\eta,t}^{\rho 0}
+
{\cal V}_{\Upsilon B^*\bar B^*,t}^{\mu\nu a}
{\cal V}_{B^*\bar B^*\eta,t}^{\rho a}.
\label{eq:SMTimeSpaceContraction}
\end{eqnarray}
The time component vanishes because
$ 
{\cal V}_{B^*\bar B^*\eta,t}^{\rho 0}
\propto
\epsilon_{0\rho 0\lambda}p_4^\lambda
=
0 .
$
Thus the leading-order contribution in the exchange kernel is purely spatial.  We then introduce
spatial indices $i,j,k$ for the $\Upsilon(2S)$, $B^*$, and $\bar B^*$
polarizations, respectively, and the propagator contributes
\begin{eqnarray}
\frac{i\delta^{ab}}{q_t^2-m_{B^*}^2+i\epsilon}
\end{eqnarray}
to the leading-order contribution.

The denominator is expanded in the near-threshold kinematics.  This is not an
instantaneous approximation: the four-momentum dependence in
$q_t^2-m_{B^*}^2$ is retained and evaluated at the leading external
kinematics, while the residual loop-momentum dependence is absorbed into
higher-order local terms.  Explicitly,
\begin{eqnarray}
q_t^2-m_{B^*}^2
=
D_{B^*}+O(|\bm l|,|\bm l|^2),
\qquad \mathrm{with} \qquad
D_{B^*}
=
m_\eta^2-2m_{B^*}E_\eta .
\label{eq:SMDBstar}
\end{eqnarray}
Additionally, a replacement for the tensor structure $T_t^{ija}$ in the $\Upsilon(2S)B^*\bar B^*$ vertex can be made, i.e., 
\begin{eqnarray}
T_t^{ija}
\longrightarrow
\delta^{ia}l^j-\delta^{ij}l^a,
\label{eq:SMTtBstarNR}
\end{eqnarray}
in which the omitted terms do not contain the loop momentum required by the
$P$-wave $B^*\bar B^*$ source, and their contributions will vanish after performing the complete loop integral.
Substituting the reduced vertices and denominator into the exchange amplitude
gives
\begin{eqnarray}
i{\cal M}_{B^*,t}
=
i\widehat C_{B^*}
\left[
l^j\epsilon^{kim}p_4^m
-
\delta^{ij}\epsilon^{kam}l^a p_4^m
\right]
\epsilon_3^{*i}\epsilon_1^j\epsilon_2^k ,
\label{eq:SMMBstar}
\end{eqnarray}
where
\begin{eqnarray}
\widehat C_{B^*}
=
\frac{
g_{\Upsilon B^*\bar B^*}
g_{B^*\bar B^*\eta}C_\eta
}{
D_{B^*}
}.
\label{eq:SMCBstar}
\end{eqnarray}
This is the contribution of $B^*$ meson exchange to the local four-point amplitude of $B^*\bar B^*\to\Upsilon(2S)\eta$.  In Eq. (\ref{eq:SMMBstar}), it can be seen that the two operator terms correspond to the ${\cal O}_4$ and
${\cal O}_1$ operator structures introduced above, respectively.  The $u$-channel diagram is obtained by
assigning the exchanged momentum as $q_u=p_1-p_4=p_3-p_2$, corresponding to
the case in which the $\eta$ is emitted from the heavy-meson line with $p_1$ rather
than from the line with $p_2$.  Comparing the resulting local four-point amplitude with the
seven-operator basis shown in Eq. (\ref{eq:SMops}), the $t$-channel $B^*$-exchange contribution generates
the ${\cal O}_1$ and ${\cal O}_4$ structures, while the $u$-channel contribution generates ${\cal O}_2$, ${\cal O}_3$, and ${\cal O}_5$.  In the subsequent $Y(10650)\to\Upsilon(2S)\eta$ loop calculation, these local
amplitudes are contracted with the $YB^*\bar B^*$ vertex, which is symmetric under the interchange of the polarization indices of the two intermediate
vector bottom mesons.  This contraction makes the $t$- and $u$-channel $B^*$-exchange contributions equal in the final decay amplitude.

\subsection{Matching to $B$-exchange amplitude}

We next consider the pseudoscalar $B$ meson exchange contribution in the same
near-threshold kinematics.  Since the exchanged particle is spinless, the
propagator does not carry Lorentz indices.  For the $t$-channel diagram, the
reduced three-point amplitudes are~\cite{Wise:1992hn,Yan:1992gz,Grinstein:1992qt,Casalbuoni:1996pg,Li:2012cs,Li:2012ss,Achasov:1994vh,Deandrea:2003pv,Colangelo:2003sa,Lin:1999ad}
\begin{eqnarray}
{\cal V}_{\Upsilon B \bar B^*,t}^{ij}
&=&
-2g_{\Upsilon B \bar B^*}
\epsilon^{ijm}
\left(
E_{\Upsilon(2S)}l^m
+
E_{B^*}p_4^m
\right),
\label{eq:SMUpsilonBBstarVertex}
\\
{\cal V}_{\eta B \bar B^*,t}^{k}
&=&
-g_{\eta B \bar B^*}C_\eta\,p_4^k .
\label{eq:SMetaBBstarVertex}
\end{eqnarray}
Here $i$, $j$, and $k$ label the $\Upsilon(2S)$, incoming $B^*$, and incoming
$\bar B^*$ polarizations, respectively.  The coupling is
$g_{\eta B \bar B^*}=2g\sqrt{m_Bm_{B^*}}/f_\pi$, and $g_{\Upsilon B \bar B^*}=\frac{m_{\Upsilon(2S)}}{f_{\Upsilon(2S)}\sqrt{m_{B}m_{B^*}}}$~\cite{Achasov:1994vh,Deandrea:2003pv,Colangelo:2003sa,Lin:1999ad} with $f_{\Upsilon(2S)}=0.497$ GeV~\cite{Huang:2018pmk}.

The second term proportional to
$E_{B^*}p_4^m$ in ${\cal V}_{\Upsilon B \bar B^*,t}^{ij}$, is independent of the loop momentum.  After multiplication by
the $P$-wave $YB^*\bar B^*$ vertex, it yields an odd integrand in $\bm l$ and therefore vanishes upon the loop integration.  The relevant part for the present decay amplitude is therefore
\begin{eqnarray}
{\cal V}_{\Upsilon B \bar B^*,t}^{ij}
\longrightarrow
-2g_{\Upsilon B \bar B^*}E_{\Upsilon(2S)}
\epsilon^{ijm}l^m .
\label{eq:SMUpsilonBBstarOdd}
\end{eqnarray}
The scalar propagator denominator is also expanded at the small loop momentum $|\bm l|$
as in the $B^*$-exchange case,
\begin{eqnarray}
q_t^2-m_B^2
=
D_B+O(|\bm l|,|\bm l|^2),
\label{eq:SMDBexp}
\end{eqnarray}
where
\begin{eqnarray}
D_B
=
m_{B^*}^2-m_B^2+m_\eta^2
-2m_{B^*}E_\eta .
\label{eq:SMDB}
\end{eqnarray}
The residual $\bm l$-dependent terms in the denominator correspond to
higher-order corrections.  Keeping the leading-order contribution, one
obtains
\begin{eqnarray}
i{\cal M}_{B,t}
=
i\widehat C_B\,
\epsilon^{ijm}l^m p_4^k\,
\epsilon_3^{*i}\epsilon_1^j\epsilon_2^k ,
\label{eq:SMMBcomponent}
\end{eqnarray}
or, equivalently,
\begin{eqnarray}
i{\cal M}_{B,t}
=
i\widehat C_B
\left[
(\bm\epsilon_3^*\times\bm\epsilon_1)\cdot\bm l
\right]
(\bm\epsilon_2\cdot\bm p_4),
\label{eq:SMMB}
\end{eqnarray}
with
\begin{eqnarray}
\widehat C_B
=
\frac{
2E_{\Upsilon(2S)}
g_{\Upsilon B \bar B^*}
g_{\eta B \bar B^*}C_\eta
}{
D_B
}.
\label{eq:SMCB}
\end{eqnarray}
Thus, it can be seen that the $t$-channel $B$-exchange amplitude contributes to the ${\cal O}_7$ operator
structure.  The $u$-channel diagram is obtained by emitting the $\eta$ from
the other heavy-meson line, with $q_u=p_1-p_4=p_3-p_2$.  It gives
\begin{eqnarray}
i{\cal M}_{B,u}
=
i\widehat C_B
\left[
(\bm\epsilon_3^*\times\bm\epsilon_2)\cdot\bm l
\right]
(\bm\epsilon_1\cdot\bm p_4),
\label{eq:SMMBu}
\end{eqnarray}
which contributes to ${\cal O}_6$.  After contraction with the $YB^*\bar B^*$
vertex, the $t$- and $u$-channel $B$-exchange diagrams give the same
contribution to the final complete loop amplitude.

Combining the $B^*$ and $B$ exchange results, the exchange-saturation
prescription completely fixes the coefficients in
Eq.~\eqref{eq:SM4pt}. With the operator convention of Eq.~\eqref{eq:SMops}, the matched low-energy coupling constants
are
\begin{eqnarray}
c_1=\widehat C_{B^*},\qquad
c_2=\widehat C_{B^*},\qquad
c_3=-2\widehat C_{B^*},
%\label{eq:SMc123}
\nonumber \\
c_4=-\widehat C_{B^*},\qquad
c_5=-\widehat C_{B^*},\qquad
c_6=\widehat C_B,\qquad
c_7=\widehat C_B .
\label{eq:SMc4567}
\end{eqnarray}
%The signs and the factor of two are fixed by the Levi-Civita ordering in the operator basis.  Thus the explicit $B$- and $B^*$-exchange amplitudes fix all seven local coefficients in terms of $\widehat C_{B^*}$ and $\widehat C_B$.

\section{Loop amplitude and partial width of $Y(10650)\to\Upsilon(2S)\eta$}

The transition amplitude of $Y(10650)\to\Upsilon(2S)\eta$ is obtained by inserting the local four-point amplitude associated with
$B^*\bar B^*\to\Upsilon(2S)\eta$ into the $B^*\bar B^*$ loop
generated by the $Y(10650)$ pole.  The $YB^*\bar B^*$ vertex is governed by
the two $1^{--}$ $P$-wave configurations relevant for the components of the pole,
\begin{eqnarray}
X^{\nu\rho}({\bm l})
&=&
i g_Y^{(1)}\delta^{\nu\rho}({\bm\epsilon}_Y\cdot{\bm l})
+i g_Y^{(5)}
\left[
\epsilon_Y^\nu l^\rho+
\epsilon_Y^\rho l^\nu
-\frac23\delta^{\nu\rho}({\bm\epsilon}_Y\cdot{\bm l})
\right].
\label{eq:Ysource}
\end{eqnarray}
The first term represents the $B^*\bar B^*(^1P_1)$ component, and the second
term represents the $B^*\bar B^*(^5P_1)$ component, and $g_Y^{(1)}=8.06$, $10.00$, $10.33$ and
$g_Y^{(5)}=4.78$, $4.98$, $4.82$ are the
$YB^*\bar B^*$ coupling constants in the $^1P_1$ and $^5P_1$ channels for the $Y(10650)$ pole scheme with
$\Lambda=0.65$, $0.75$, and $0.85~\mathrm{GeV}$, respectively.  Both terms are symmetric
under $\nu\leftrightarrow\rho$, where $\nu$ and $\rho$ denote the polarization
indices of the two intermediate vector bottom mesons.

The loop integral contains two factors of the relative momentum $\bm l$: one
from the $P$-wave $YB^*\bar B^*$ vertex and the other from the local four-point
transition amplitude of $B^*\bar B^*\to\Upsilon(2S)\eta$.  Rotational invariance reduces the tensor integral to
\begin{eqnarray}
\int^{\alpha_p}\frac{d^3{\bm l}}{(2\pi)^3}
\frac{l^r l^m}
{E_Y-2m_{B^*}-{\bm l}^{2}/(2\mu_{B^*B^*})+i0}
=
\frac{\delta^{rm}}{3}I_2(E_Y,\alpha_p),
\label{eq:SMavg}
\end{eqnarray}
which defines
\begin{eqnarray}
I_2(E_Y,\alpha_p)
=
\int_{|{\bm l}|<\alpha_p}
\frac{d^3{\bm l}}{(2\pi)^3}
\frac{{\bm l}^{2}}
{E_Y-2m_{B^*}-{\bm l}^{2}/(2\mu_{B^*B^*})+i0}.
\label{eq:SMI2}
\end{eqnarray}
Here, $\alpha_p$ restricts the loop momentum to the near-threshold region and
is varied from $1.0$ to $1.3~{\rm GeV}$ in the numerical analysis.

After contracting the polarization tensors in the final decay amplitude, the $B^*$-exchange contribution gives
\begin{eqnarray}
{\cal A}^{\rm full}_{B^*}
&=&
-\frac{1}{4m_{B^*}^2}I_2(m_Y,\alpha_p)2\widehat C_{B^*}
\left(
\frac43g_Y^{(1)}
+
\frac{10}{9}g_Y^{(5)}
\right)
({\bm\epsilon}_Y\times{\bm\epsilon}_\Upsilon^*)\cdot{\bm p}_\eta .
\label{eq:SMABstar}
\end{eqnarray}
The pseudoscalar $B$-exchange part contributes
\begin{eqnarray}
{\cal A}^{\rm full}_{B}
&=&
-\frac{1}{4m_{B^*}^2}I_2(m_Y,\alpha_p)2\widehat C_{B}
\left(
\frac23g_Y^{(1)}
-
\frac{10}{9}g_Y^{(5)}
\right)
({\bm\epsilon}_Y\times{\bm\epsilon}_\Upsilon^*)\cdot{\bm p}_\eta .
\label{eq:SMAB}
\end{eqnarray}
The full amplitude can therefore be written as
\begin{eqnarray}
{\cal A}^{\rm tot}
&=&
{\cal N}^{\rm tot}
({\bm\epsilon}_Y\times{\bm\epsilon}_\Upsilon^*)\cdot{\bm p}_\eta ,
\nonumber\\
{\cal N}^{\rm tot}
&=&
-\frac{1}{2m_{B^*}^2}I_2(m_Y,\alpha_p)
\left\{
\widehat C_{B^*}
\left[
\frac43g_Y^{(1)}
+
\frac{10}{9}g_Y^{(5)}
\right]
+
\widehat C_{B}
\left[
\frac23g_Y^{(1)}
-
\frac{10}{9}g_Y^{(5)}
\right]
\right\}.
\label{eq:SMAtot}
\end{eqnarray}
The partial width then follows as
\begin{eqnarray}
\Gamma[Y(10650)\to\Upsilon(2S)\eta]
=
\frac{|{\bm p}_\eta|^3}{12\pi M_Y^2}
\left|{\cal N}^{\rm tot}\right|^2 .
\label{eq:SMwidth}
\end{eqnarray}

% %otherwise go to head of supplemental and generate the SM as a single file with indepedent reference. 
%%latexmkrc is used for set up the crossref between main tex and the supplemental.
% %note for the SM, the heperlink from crossref from the main is not correct.

%\appendix
%\clearpage
%\onecolumngrid

%\bibliography{refs}
\end{document}